\documentstyle[aps,prl,twocolumn,epsf]{revtex}
\begin{document}

\title{ Scaling Properties of the Giant Dipole Resonance Width in Hot
Rotating nuclei}

\author{Dimitri Kusnezov$^1$
     \footnote{E--mail: dimitri@nst.physics.yale.edu},
        Y. Alhassid$^1$
     \footnote{E--mail: yoram@nst.physics.yale.edu}, and
        K. A. Snover$^2$
     \footnote{E--mail: snover@npl.washington.edu}}
\address{$^1$Center for Theoretical Physics,
              Sloane Physics Laboratory,
              Yale University, New Haven, CT 06520-8120 \\
         $^2$Nuclear Physics Laboratory, Box 354290,
              University of Washington, Seattle, WA 98195}
\date{\today}
\maketitle
\begin{abstract}
We study the systematics of the giant dipole resonance
 width  $\Gamma$ in hot rotating nuclei as a function of  temperature
 $T$, spin  $J$  and mass $A$.   We compare available experimental
results with theoretical calculations that include thermal shape fluctuations
 in nuclei  ranging  from $A=45$ to  $A=208$.    Using the appropriate scaled
 variables, we   find a
simple phenomenological function  $\Gamma(A,T,J)$ which approximates
 the global behavior of the giant dipole resonance width in the
liquid drop model. We reanalyze recent experimental and
theoretical results for  the resonance width in Sn isotopes
 and $^{208}$Pb.

\end{abstract}
\draft\pacs{PACS numbers: 05.45.+b, 02.20.-a, 21.60.Fw, 03.65.-w}

\narrowtext

Hot rotating nuclei are usually produced in heavy ion fusion reactions
 through  transfer of the energy and angular momentum of the relative
 motion into internal degrees of freedom. The
resulting hot nucleus can decay through particle and gamma-ray emission.
 From the decay patterns of these nuclei one can hope to understand their
properties under extreme conditions such as high temperature and spin.
A particularly useful experimental probe in the study of hot nuclei
 has been the giant dipole resonance (GDR) \cite{S86,G92}.  At zero
temperature,
the GDR vibrational frequency is inversely proportional to the length of
 the axis along which the vibration occurs, and  the quadrupole deformation
 of the nucleus can be inferred from the splitting of the GDR peak.
At finite temperature the nuclear shape fluctuates, and the relationship
between the shape and the observed resonance properties is more complex.
In the adiabatic limit the observed GDR strength function is calculated
through an average over a thermal ensemble of shapes corresponding
 to all quadrupole degrees of freedom \cite{A92}. These include both
 the intrinsic shape and the
 nuclear orientation with respect to its rotation axis.  The 
fluctuation theory explains
 successfully both  the observed cross-section and angular anisotropy
 of the GDR radiation \cite{AB90,eric,gund}.

 In recent  years, a wealth of experimental  results for the GDR has
 become available in wider regions of 
temperature and spin \cite{gund,exp,exp1,R96}.  In 
the fusion experiments,
 higher excitation energies are usually accompanied by larger
 amounts of angular momentum transfer.  However, in recent  inelastic
 scattering  experiments of light particles (e.g. alpha particles) from heavy 
nuclei the GDR could be excited over a range of temperatures
without substantial angular momentum transfer \cite{R96}.
  Although detailed theoretical analyses of the GDR have been done
in many nuclei, a  comprehensive study of its global features
 has been lacking.
In this letter we present  a systematic analysis of the GDR width
 as a function of temperature $T$, spin $J$ and mass $A$. We
compare available experimental results with theoretical calculations in  nuclei
ranging  from $A \sim 45$ to  $A \sim 208$.  The calculations include thermal 
 shape fluctuations using both  Nilsson-Strutinsky (NS) and liquid drop (LD) 
free energy surfaces.
We find that  by introducing appropriate scaling of the variables
  it is possible to approximate  the GDR width
$\Gamma(A,T,J)$ in the LD regime by a simple phenomenological function.

A theory of hot rotating nuclei was developed in the framework of the
 Landau theory, where the quadrupole deformation parameters in the
laboratory frame $\alpha_{2\mu}$  ($\mu =-2,\ldots,2$) play the role of
 the order parameters \cite{ALZ87}.  The free energy at constant temperature $T$
 and angular velocity $\omega$ is expanded in the form
\begin{eqnarray}\label{free_omeg}
 F(T,\omega; \alpha_{2\mu})=F(T,\omega=0;\beta,\gamma) -
   \frac{1}{2} \left(\hat\omega \cdot {\cal I}\cdot {\hat\omega}\right)
\omega^2 \;,
\end{eqnarray}
where $\beta,\gamma$ are the intrinsic shape parameters.
 The quantity  $\hat\omega \cdot {\cal I}\cdot {\hat\omega} =
 I_{x'x'}\sin^2\theta\cos^2\phi  +
I_{y'y'}\sin^2\theta\sin^2\phi + I_{z'z'}\cos^2\theta$ is the moment
of inertia about the rotation axis $\hat\omega$, expressed in terms of the
principal moments of inertia $I_{x'x'},I_{y'y'},I_{z'z'}$ and the Euler angles
$\Omega=(\psi,\theta,\phi)$ that describe the nuclear orientation with
 respect to the rotation axis.
 In the finite nuclear system, fluctuations in the order parameters
are important and the probability of finding the nucleus  in a state
 with deformation $\alpha_{2 \mu}$ is given by the Boltzmann
factor $\exp [-F(T,\omega; \alpha_{2\mu})/T]$ \cite{LA84,GD85,AB88,PYB88}.
At high spins it is
necessary to project on constant spin \cite{AW93}, and  in the saddle point
approximation the free energy at spin $J$ is the
Legendre transform of  (\ref{free_omeg})
\begin{eqnarray}\label{free_J}
 F(T,J; \alpha_{2\mu})=F(T,\omega=0;\beta,\gamma) +
      \frac{(J+ 1/2)^2}{2{\hat\omega}\cdot {\cal I}\cdot {\hat\omega}}
\;.
\end{eqnarray}

In the non-rotating ($\omega=0$) case, the GDR absorption cross-section
 at a fixed shape $\alpha_{2\mu}$  is described by a superposition of
Lorentzians  with centroids inversely proportional
 to the lengths $R_j$ of the corresponding principal axes:   $E_j=E_0(R_0/R_j)$,
 and widths satisfying a power law $\Gamma_j=\Gamma_0(A)(E_j/E_0)^\delta$ (with
 $\delta=1.6$).    The $R_j$ depend
 on the intrinsic shape through the Hill-Wheeler parametrization
$R_j=R_0 \exp \left[-\sqrt{5/4\pi} \beta \cos(\gamma - 2\pi j/3) \right]$.
$E_0$ and $\Gamma_0(A)$ are the mass-dependent
energy and width, respectively, for a spherical shape
and are assumed to be temperature independent.  For $\omega \neq 0$, 
 the eigenfrequencies $E_j(\omega)$ are affected by  the Coriolis force,
 and it is necessary to transform from
 the rotating frame to the laboratory frame \cite{AB88}.

  At constant temperature and spin, the observed GDR cross-section is
calculated  by a  thermal average of the shape-dependent  cross-section
\cite{AW93}
\begin{equation}\label{fluct}
 \sigma(E_\gamma;J,T)={1 \over Z} \int  {\cal D}[\alpha]
 { e^{-F(T,J;\alpha_{2\mu})/T}  \over (\hat\omega \cdot{\cal
      I}\cdot\hat\omega)^{3/2} }  \sigma(E_\gamma;\omega,\alpha_{2\mu})
\end{equation}
where the measure is given by \cite{AB88}
${\cal D}[\alpha]=\beta^4\mid\sin 3\gamma\mid d\beta d\gamma d\Omega$,
and $Z=\int {\cal D}[\alpha] \exp[-F/T]/(\hat\omega
\cdot{\cal I}\cdot\hat\omega)^{3/2}$ is the partition function.

The free energy surface at $\omega=0$ and the principal moments of inertia
 are calculated  using  either an NS approach (which includes shell 
corrections), or  the LD model.  At  higher temperatures,
 shell  effects melt and both approaches agree well with each other.

  We have carried out a comprehensive study of  the
  GDR over a wide  range of nuclei for which experimental data
 exist \cite{exp}:   $^{45}$Sc, $^{59,63}$Cu,
$^{90}$Zr, $^{92,100}$Mo, $^{106-120}$Sn, $^{156}$Dy, $^{166,168}$Er, 
$^{188}$W,  and $^{208}$Pb. We find that a
 simple behavior emerges in the LD model.
 We first examine the spin dependence of the width
$\Gamma$ at fixed temperature $T$.  Fig  \ref{fig1}(a)  shows  $\Gamma$
versus spin for $^{59,63}$Cu. The symbols are the experimental results
 and the lines are  theoretical LD calculations.
The overall agreement between theory
and experiment is good (except at low T  where the LD model does not
apply); the width is insensitive to spin for $J \alt 20 \hbar$
 and increases at higher spins. Similarly,  in  \ref{fig1}(b) we show the spin
 dependence of width in $^{106}$Sn, where our  LD calculations 
at $T=1.8$ MeV (solid) reproduce well the experimental behavior\cite{exp1},
 and account for up to  $\sim 20\%$ enhancement at high spins over the
 calculations of  \cite{exp1} (dashes).  For $^{106}$Sn the
width remains insensitive to spin up to a higher spin of $J \alt 30 \hbar$. 
In Fig. \ref{fig1}(c) we show
 $\Gamma$ versus $J$ at $T=2$ MeV  for several nuclei
in different mass regions for spins up to their respective fission limit. 
 The sensitivity of the GDR width to spin is larger  for the lighter
nuclei as is expected from their smaller moment of inertia.
We have investigated several possible scalings to relate the
 reduced widths $\Gamma(T,J,A)/\Gamma(T,J=0,A)$ of various
 masses.  At high spins, the rotational energy $J^2/2I$ dominates.
Since for a rigid body $I \propto A^{5/3}$, this suggests a scaling
 of the spin by $A^{5/6}$.  Fig. \ref{fig1}(d-e)  shows the reduced width
as a function of $\xi \equiv J/A^{5/6}$.  At a fixed temperature the
reduced width for various masses falls approximately on a single
curve. While this is clearly not an exact scaling of the theory described by 
Eqs. (\ref{fluct}) and (\ref{free_J}),
it provides a rather good approximation.  The  scaling improves with
increasing temperature.
We remark that  a significant mass dependence of the width is
 observed when plotted either versus angular velocity $\omega$
 or  the rotation parameter $y$ of the LD \cite{cps}.

The scaling curves in Fig. \ref{fig1}(e) exhibit
a significant temperature dependence. We choose the reduced width
$\Gamma(T_0,J,A)/\Gamma(T_0,J=0,A)$ at  $T_0=1$MeV  to be
 our `reference'  function $L(\xi)$  with $\xi= J/A^{5/6}$
The reduced widths at different
temperatures are related through the power law
$[\Gamma(T,J,A)/\Gamma(T,J=0,A)]^{(T/T_0+3)/4}$ as is shown in Fig. 
\ref{fig1}(f).
Hence  the approximate  spin dependence of $\Gamma$ is described by
$\Gamma(T,J,A)/\Gamma(T,J=0,A) \approx   \left[L(\xi)\right]^{4/(T/T_0+3)}$.

Next we  examine the temperature (and mass) dependence of the
 width at zero spin $\Gamma(T,J=0,A)$.  In the left panel of Fig. \ref{fig2} we
show the experimental width at low spin ($J \alt 20 \hbar$) for $^{59,63}$Cu
 as a function of $T$ in comparison with exact LD calculations
(solid line). The quantity $ \Gamma(T,J=0,A)-\Gamma_0(A)$
 (where $\Gamma_0(A)$ is the width for a spherical shape) increases
  monotonically from zero as a function of $T$.  At high
 temperatures it  behaves as  $\sqrt{T}$:  using the leading order term
$B \beta^2$ for the LD free energy (with $B$ constant), we
 can remove the temperature dependence in the
Boltzmann factor $\exp[-B \beta^2/T]$, by  scaling $\beta$ by $\sqrt{T}$.
  This works well only for temperatures  $T\agt 2$ MeV,
and a much better global fit is obtained from
$\Gamma(T,J=0,A)-\Gamma_0(A) \approx c(A) \log(1+T/T_0)$,
where $T_0=1$ MeV is the reference temperature and $c(A)$ is a
 constant  depending weakly on $A$. In the right panel
of Fig. \ref{fig2},  we show this fitting function
 for $^{90}$Zr  (solid line) and compare it with the liquid drop
calculations (squares).
The dashed line demonstrates the $\sqrt{T}$  behavior at large $T$.
The function $c(A)$ depends on the the choice of $\Gamma_0$,
 since increasing the width
$\Gamma_0$ does not result in a constant  shift of the width at all
temperatures, but rather a modification of the prefactor $c(A)$.
A parametrization which seems to work well over the mass range studied
(and for our physical choices of $\Gamma_0$)  is
$ c(A) \approx 6.45 - A/100$. 
% For the nuclei studied, only  $^{90}$Zr
%has a noticeable deviation of its fitted  $c(A)$  from this formula.

We conclude that a good phenomenological formula to describe the
 global dependence of the LD GDR width on
 temperature, spin  and mass is:
\begin{eqnarray}\label{width}
  \Gamma(T,J,A) &=& \Gamma(T,J=0,A)
     \left[L\left(\frac{J}{A^{5/6}}\right)\right]^{4/[(T/T_0)+3]}\\ \nonumber
  \Gamma(T,J=0,A)  &=& \Gamma_0(A) + c(A)\log ( 1 + T/T_0) \;.
\end{eqnarray}
 $\Gamma_0(A)$ is  usually extracted from the measured ground
state GDR, and $T_0=1$ MeV is  a reference temperature.  
 $L(\xi)$ is the scaling function shown in Fig.\ \ref{fig1} (f),  which can be
 approximately fitted by $L(\xi) - 1\approx  
1.8\left[ 1 + e^{(1.3- \xi)/0.2} \right]^{-1}$.  Eqs.  (\ref{width})
provide an approximate description of the systematic behavior of the GDR width
in nuclei where the LD model is valid, i.e. in nuclei where shell
effects are small or at temperatures where shell effects have already melted.
In the top panel of Fig. \ref{fig3} we correlate the theoretical estimates 
based on  (\ref{width}) with known experimental results
\cite{gund,exp,exp1,R96}.  In the bottom panel of Fig. \ref{fig3}(b) we show 
the ratio between the experimental width $\Gamma_{exp}(T,J,A)$ and the
`theoretical' width $\Gamma(T,J=0,A)$ calculated from (\ref{width}) as a
function of  $\xi$.

 The scaling function $L(\xi)$ is seen to be essentially constant for
 $\xi\alt 0.6$
 (indicated by dashed line in Fig. \ref{fig1}(f)).
 Thus the width is approximately spin-independent up to a
spin of $J_1\sim 0.6 A^{5/6}$.
  It is interesting to compare $J_1$ with  the maximal
angular momenta $J_{max}(A)$  for which the fission barrier height  is
 still  larger  than $\sim 8$ MeV, guaranteeing reasonable stability 
against fission \cite{cps}.  We find that for nuclei with $A\agt 200$, 
$J_{max}\leq J_1$, and there is no significant spin dependence of the 
GDR width (see e.g. $^{208}$Pb in Fig. \ref{fig1}(c)).

 Shell corrections can play a role at lower temperatures.  Here
 we focus on two nuclei of recent experimental \cite{R96}
 and theoretical \cite{OBB96} interest, $^{120}$Sn and $^{208}$Pb.
 Fig.  \ref{fig4} shows the results of our calculations of the width as a 
function of temperature using both LD (dotted line) and NS (solid line) free 
energy surfaces.  We have used  $\Gamma_0= 3.8$ MeV\cite{AB88}
 for  $^{120}$Sn and $^{208}$Pb. Our results are compared
 with the recent  calculations of Ref. \cite{OBB96}, also shown in
 Fig. \ref{fig4}  (dashes and dot-dashes)\cite{Gu}.
 For $^{208}$Pb our calculated widths at temperatures above 1 MeV
 are significantly larger than those of Ref.  \cite{OBB96}.
Similarly for $^{120}$Sn,  our calculated widths  are  larger
 than those of Ref.  \cite{OBB96}  at large temperatures
even though our  assumed $\Gamma_0$
(3.8 MeV) is smaller than the one used  in Ref. \cite{OBB96} (5 MeV).
 When compared with  our newly calculated widths, the
experimental results of Refs. \cite{R96}  (open diamonds in Fig. \ref{fig4})
 show significant deviations.
We have  re-evaluated the temperatures corresponding to the
$^{120}$Sn and $^{208}$Pb inelastic scattering data, and found
new temperatures (solid diamonds in Fig. \ref{fig4}) that are substantially 
smaller than the values quoted
in Refs. \cite{R96}. These revised data points are  in better agreement
 with our calculations (except for the two highest temperature points in Sn).
 For  $^{208}$Pb they are also in better agreement 
 with the fusion data (shown by x's). With the above revision of  both
 theory and experiment,  the new results confirm the
conclusions of  Ref. \cite{OBB96}:  shell effects on the GDR width are 
negligible in Sn, while the large
shell corrections in Pb cause a suppression of the width at low temperatures.
Similar suppression of the width due to shell effects can be seen in the
 calculations for  $^{140}$Ce in Ref.\  \cite{AB88}.
In revising the experimental temperatures, we have included the effect of
energy lost by particle evaporation prior to $\gamma$-decay \cite{S86} in both
$^{120}$Sn and $^{208}$Pb by using the computer code Cascade\cite{P77} to
average over the decay cascades\cite{kurt}.  
In addition, in $^{208}$Pb we included the
effect of the strong shell correction on the temperature, 
%and in $^{120}$Sn  we
%assumed a constant level density parameter \cite{S98}
%$a = A/9$ consistent with Ref.
%\cite{R81} and with experiment \cite{N94}. 
and in $^{120}Sn$ we assumed the Reisdorf\cite{R81} level density, which has
a small shell correction and a nearly constant level density parameter\cite{S98}
$a\approx A/9$ consistent with experiment\cite{N94}.
All of these corrections 
 reduce the temperatures quoted in Refs.\cite{R96}.

In conclusion, we have studied the systematics of  the GDR width in
 hot rotating nuclei over a broad range of  nuclear masses  in the framework of
 the thermal fluctuation theory.  In the liquid drop limit we have
 found a phenomenological formula that describes well  the 
width behavior as a function of temperature, spin and mass.

This work was supported in part by DOE grants DE-FG02-91ER40608 and
DE-FG-97ER41020.

\begin{figure}
%\epsfxsize=8 cm
%\centerline{\epsffile{fig1.ps}}
\vspace{2 mm}
\caption{Spin dependence of the GDR width $\Gamma$ (left column top to bottom
are (a)-(c) and right (d)-(f)). (a) Comparison between experiment
(symbols) and LD theory at $T=1.8$ (solid line) and $T=1.5, 2.1$ MeV
 (lower and upper dotted lines, respectively)
in $^{59,63}$Cu.  (b) The experimental widths in $^{106}$Sn 
 \protect\cite{exp1} compared
with our LD widths (solid line) and those of \protect\cite{exp1} (dashes). 
 (c) Systematics
of $\Gamma$ versus spin $J$ (using LD surfaces) at $T=2$ MeV for $^{59}$Cu,
$^{90}$Zr, $^{120}$Sn, and $^{208}$Pb.  (d)   $\Gamma(J,T,A)/\Gamma(T,J=0,A)$
versus $\xi \equiv J/A^{5/6}$ for all nuclei shown in (c) and for $T=2$ MeV.
(e)  Same as (d) but for $T=1,4$ MeV displaying temperature dependence at 
$\xi>1$. (f) $[\Gamma(T,J,A)/\Gamma(T,J=0,A)]^{(T/T_0+3)/4}$ versus $\xi$ for
$T=1,2,3,4$ MeV for the nuclei in (c).  The
solid curve is the scaling function $L(\xi)$ (see text).}
\label{fig1}

%\epsfxsize=8 cm
%\centerline{\epsffile{fig2.ps}}
\vspace{2 mm}
\caption{Temperature dependence of the width.
 Left: $\Gamma$ (for $J \protect\alt 20 \hbar$)
 as a function of $T$ for $^{59,63}$Cu from experiment (symbols) and
 theory (solid line).  Right: $\Gamma(T,J=0,A) - \Gamma_0$ as a function
 of $T$ for $^{90}$Zr from the LD calculations (boxes),  a 
 fit to  $c(A) \log(1 + T/T_0)$ (solid line) and a
 $\protect\sqrt{T}$ behavior (dotted line) which generally fits well at large 
$T$.}
\label{fig2}

%\epsfxsize=8 cm
%\centerline{\epsffile{fig3.ps}}
\vspace{2 mm}
\caption{ Comparison of  experimental widths with the
phenomenological width formula (\protect\ref{width}). Top:
Experimental $\Gamma$ vs. theoretical scaled $\Gamma$ for selected nuclei in
the mass range  $A \sim 45$ to  $208$. Bottom: Ratio  of experimental
$\Gamma$ to theoretical scaled $\Gamma(T,J=0,A)$ versus $\xi=J/A^{5/6}$. The 
solid line is the scaling function $L(\xi)$. }
\label{fig3}

%\epsfxsize=8 cm
%\centerline{\epsffile{fig4.ps}}
\vspace{2 mm}
\caption{ Temperature dependence of the GDR width in $^{120}$Sn (left)
 and $^{208}$Pb (right).  Top: our calculated widths using
NS (solid) and LD (dots) are compared with
similar lines calculated in Ref. \protect\cite{OBB96} (dashes are LD 
and dot-dashes are NS), and with the experimental
results of Refs. \protect\cite{R96} (open diamonds).
Our calculations are the curves
giving larger widths at higher temperatures.  Bottom: Our theoretical curves
compared with the revised data points (solid diamonds).
 In addition, fusion evaporation data is included (crosses)\protect\cite{exp}.}
\label{fig4}

\end{figure}


\begin{thebibliography}{99}
\bibitem{S86} K.A.  Snover,  Ann. Rev. Nucl. Part. Sci.  {\bf 36}, 545 (1986).
\bibitem{G92} J. Gaardhoje,  Ann. Rev. Nucl. Part. Sci.  {\bf 42}, 483 (1992).
\bibitem{A92} Y. Alhassid, in {\it New Trends in Nuclear Collective Dynamics},
 p. 41,  Y. Abe, H. Horiuchi and K. Matsuyanagi, eds.,  Springer Verlag,
 New York, 1992.
\bibitem{AB90} Y. Alhassid and B. Bush, Phys. Rev. Lett. {\bf 65}, 2527 (1990).
\bibitem{eric} W.E. Ormand et al, Phys. Rev. Lett. {\bf 69}, 2905 (1992).
\bibitem{gund} J.H. Gundlach, et al,  Phys. Rev. Lett. {\bf 65},  2523 (1990).
\bibitem{exp}  J.J. Gaardhoje et al, Phys. Rev. Lett. {\bf  53}, 148 (1984);
C.A. Gossett et al, Phys. Rev. Lett. {\bf 54}, 1456 (1985);
J. J. Gaardhoje et al, Phys. Rev. Lett. {\bf  56}, 1783 (1986);
M. Kicinska-Habior  et al, Phys. Rev. C {\bf  36}, 612 (1987);
D.R. Chakrabarty et al, Phys. Rev. {\bf C 36}, 1886 (1987);
A Bracco  et al, Phys. Rev. Lett. {\bf 62}, 2080 (1989);
 K.A. Snover, AIP Conf. Proc. {\bf  259}, 299 (1992);
M. Kicinska-Habior et al, Phys. Rev. C {\bf 45} 569 (1992);
M. Kicinska-Habior et al, Phys. Lett. B {\bf 308} 225 (1993);
J.P.S. van Schagen  et al, Phys. Lett.  B {\bf 343}, 64 (1995);
A. Bracco  et al, Phys. Rev. Lett.  {\bf 74}, 3748 (1995);
Z.M. Drebi et al, Phys. Rev. C {\bf 52},  578 (1995);
R. Butsch et al,  Phys. Rev. C {\bf 41},  1530 (1990);
D.R. Chakrabarty et al, Phys. Rev. {\bf C 53}, 2739 (1996).
\bibitem{exp1} M. Mattiuzzi et al, Nucl. Phys. {\bf A612}, 262 (1997).
\bibitem{R96} E. Ramakrishnan et al., Phys. Rev. Lett. 76, 2025(1996):
 E. Ramakrishnan et al., Phys. Lett. B383, 252(1996).
\bibitem{ALZ87} Y. Alhassid, J. Zingman, and S. Levit, Nucl. Phys.
 {\bf A 469}, 205 (1987).
\bibitem{LA84} S. Levit and Y. Alhassid, Nucl. Phys. {\bf A413}, 439 (1984).
\bibitem{GD85} M. Gallardo et al, 
% M. Diebel, T. Dossing and R.A. Broglia,
     Nucl. Phys. {\bf A443}, 415 (1985).
\bibitem{AB88} Y. Alhassid, B. Bush and S. Levit, Phys. Rev. Lett. {\bf 61},
1926 (1988);
 Y. Alhassid and B. Bush,  Nucl. Phys.  {\bf A 509}, 461 (1990).
\bibitem{PYB88}  J.M. Pacheco, C. Yannouleas, and R.A. Broglia,
  Phys. Rev. Lett. {\bf 61}, 294 (1988).
\bibitem{AW93}  Y. Alhassid and N. Whelan, Nucl. Phys. {\bf A 565}, 427 (1993).
\bibitem{cps} S. Cohen, F. Plasil and W. Swiatecki,  Ann. Phys. (NY)
{\bf 82}, 557 (1974).
\bibitem{OBB96}  W.E. Ormand, P.F. Bortignon, and R.A. Broglia,  Phys. Rev.
Lett.  {\bf 77},  607 (1996);  W.E. Ormand et al, Nucl. Phys. {\bf A 614},
217 (1997).
\bibitem{Gu} We use the empirical LD parameters of [16] while [17] uses that of
Guet et al, Phys. Lett. {\bf B205} (1988) 427.
\bibitem{P77} F. Puhlhofer, Nucl. Phys.  {\bf A 280}, 267(1977).
\bibitem{kurt} Similar $T$ corrections have been computed for other
fusion--evaporation data where necessary.
\bibitem{R81} W. Reisdorf, Z. Phys. {\bf A300}, 227(1981).
\bibitem{S98}  We checked that the use of our improved level density in the
interpretation of the $^{120}$Sn $\gamma$-ray spectra should not affect
substantially the extracted GDR widths.
\bibitem{N94} G. Nebbia et al., Nucl. Phys. {\bf A 578}, 285(1994).
\end{thebibliography}
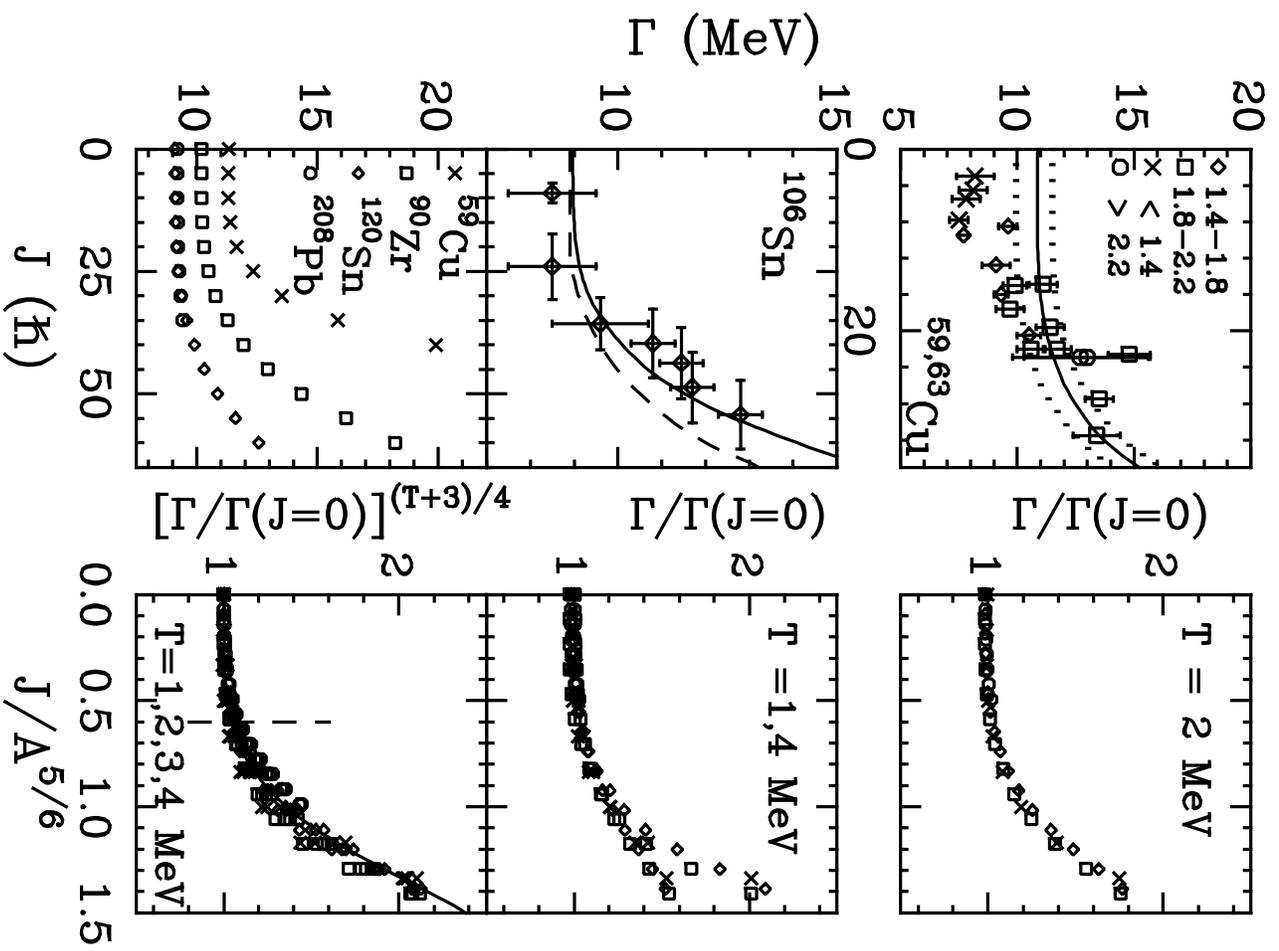
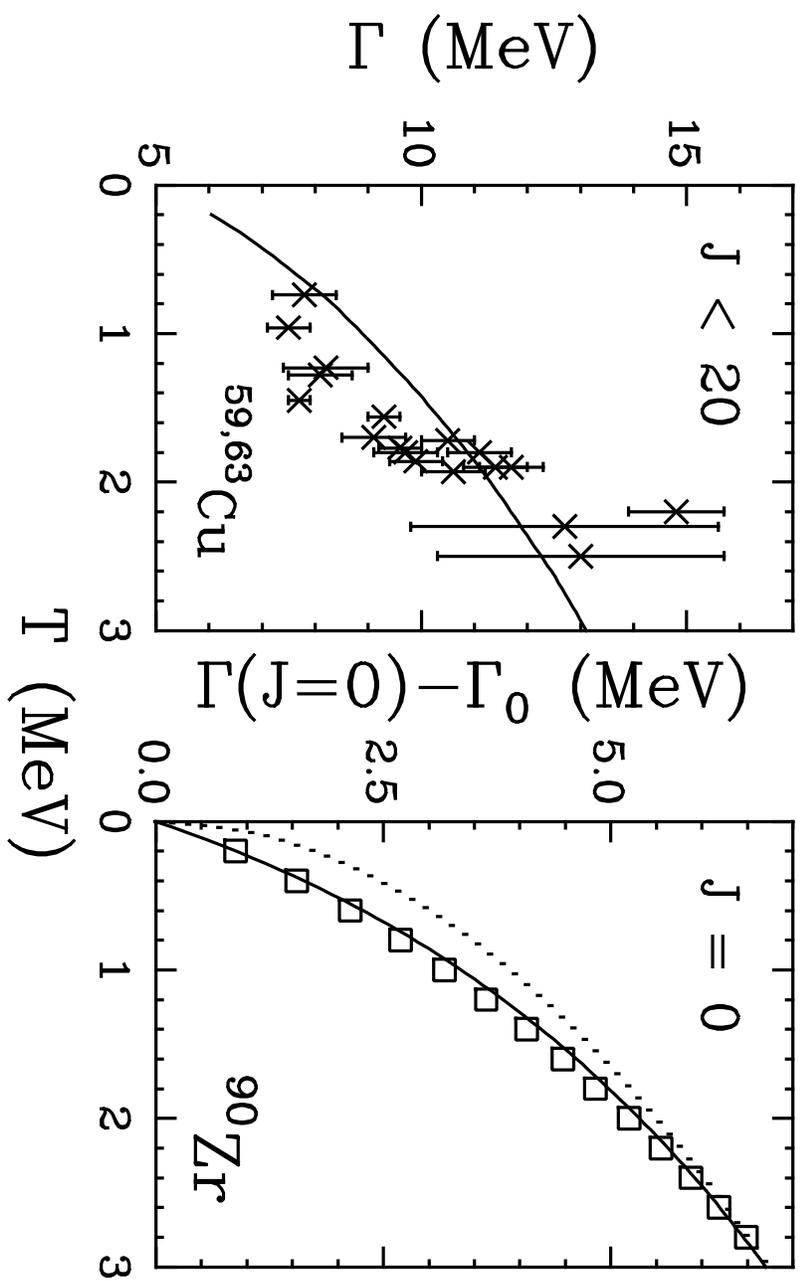
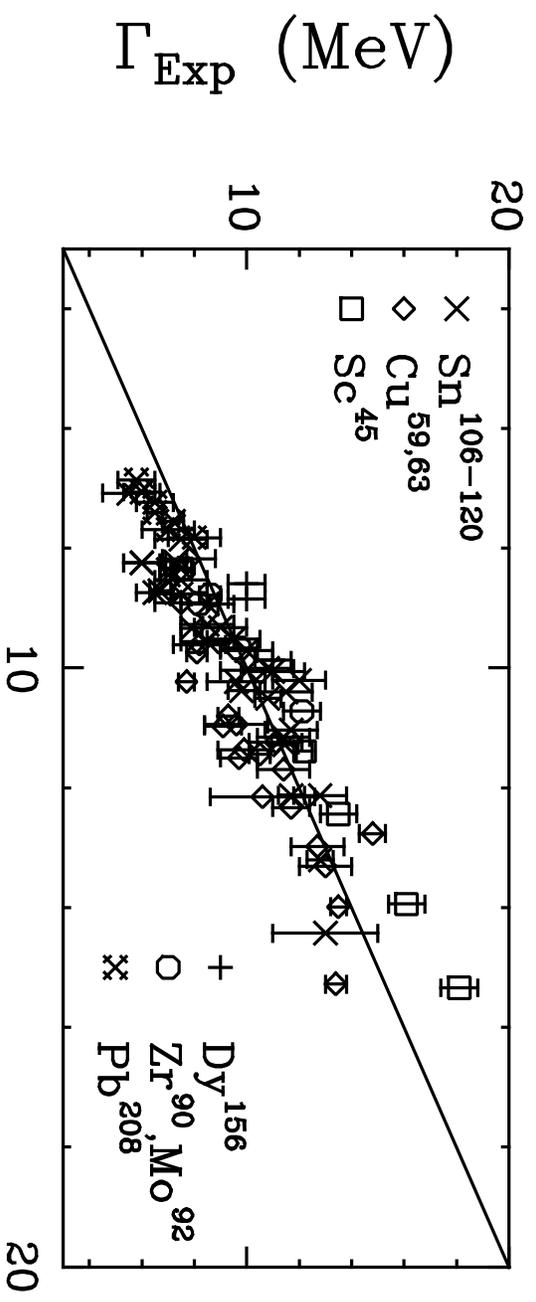
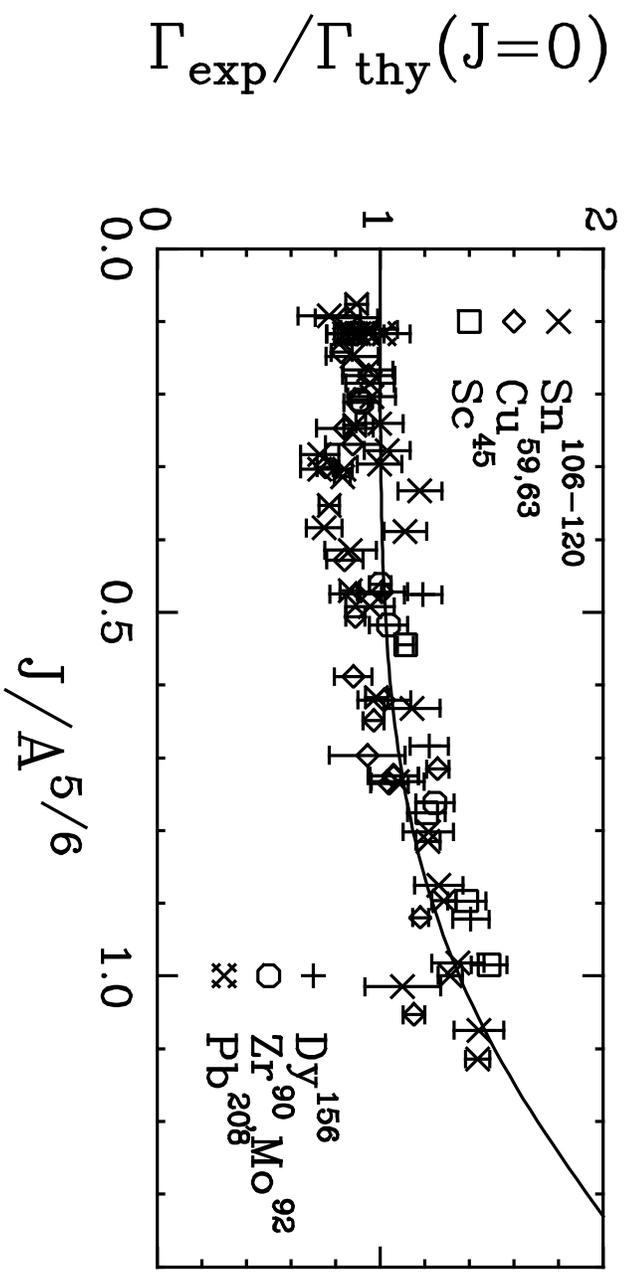
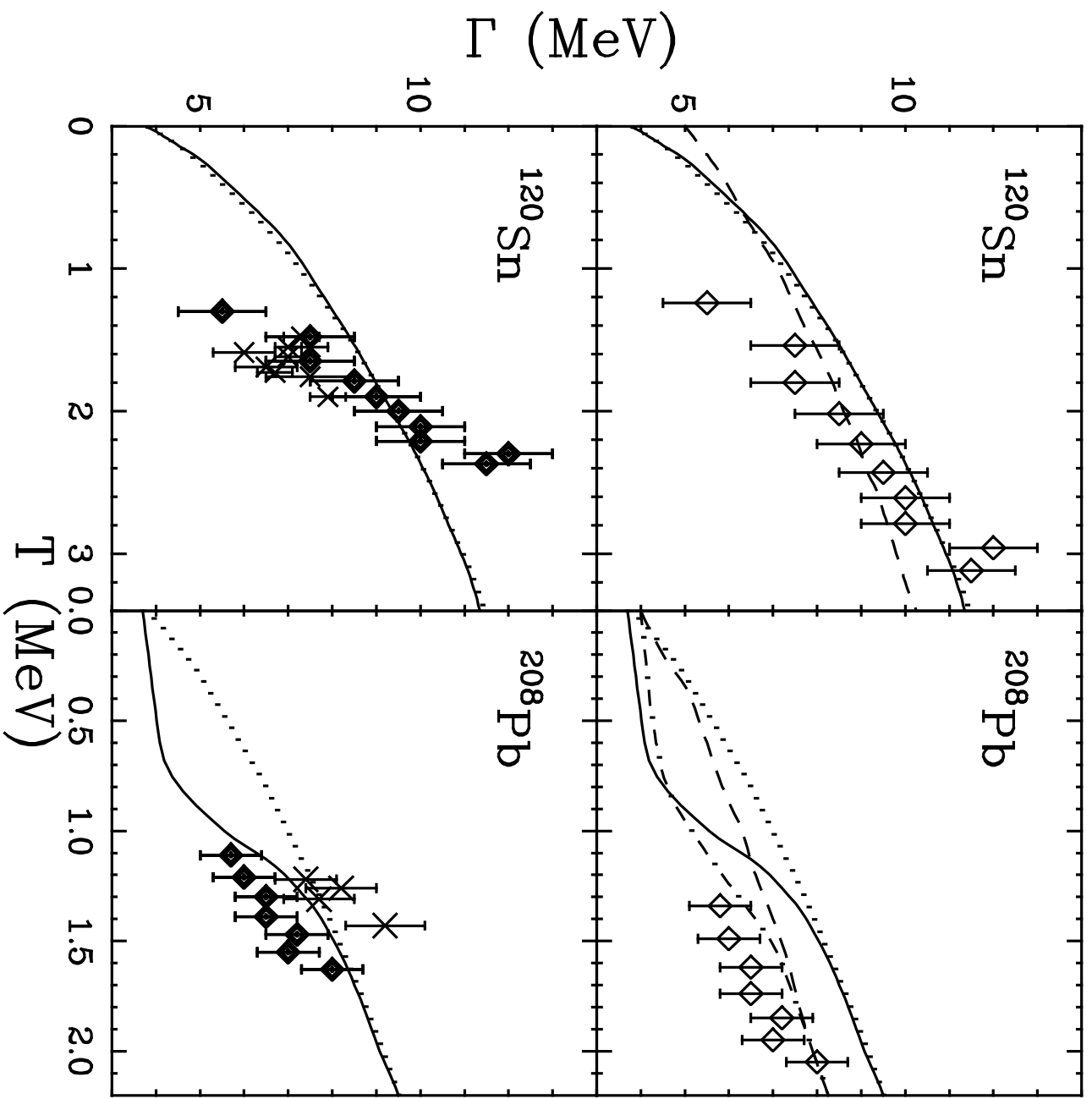
\end{document}